\documentstyle[11pt,aaspp4,flushrt,tighten]{article}
\singlespace
\newcommand{\cmm}{\,{\rm cm}^{-2}}

\newcommand{\Lya}{Ly$\alpha\ $}
\newcommand{\Lyb}{Ly$\beta\ $}
\newcommand{\lya}{Ly$\alpha\ $}

\newcommand{\etal}{et~al.\ }
\def\kms{\,{\rm km\,s^{-1}}}
\def\kmsmpc{\,{\rm km\,s^{-1}\,Mpc^{-1}}}
\def\msun{\,{\rm M_\odot}}

\def\spose#1{\hbox to 0pt{#1\hss}}
\def\lta{\mathrel{\spose{\lower 3pt\hbox{$\mathchar"218$}} \raise 2.0pt\hbox{$\mathchar"13C$}}}
\def\gta{\mathrel{\spose{\lower 3pt\hbox{$\mathchar"218$}} \raise 2.0pt\hbox{$\mathchar"13E$}}}

\def\rei{{\rm rei}}
\def\obs{{\rm obs}}
\def\rec{{\rm rec}}
\def\em{{\rm em}}
\def\b{{\rm blue}}
\def\r{{\rm red}}
\def\HI{\hbox{H~$\scriptstyle\rm I\ $}}
\def\bHII{\hbox{\bf H~$\scriptstyle\bf II\ $}}
\def\HII{\hbox{H~$\scriptstyle\rm II\ $}}

\def\nHI{{\rm HI}}
\def\nH{{\rm H}}

\newcommand\beq{\begin{equation}}
\newcommand\eeq{\end{equation}}

\begin{document}

\title{The Earliest Luminous Sources and the Damping Wing of the Gunn-Peterson 
Trough}
\author{Piero Madau\altaffilmark{1,2,3} and Martin J. Rees\altaffilmark{1}}

\altaffiltext{1}{Institute of Astronomy, Madingley Road, Cambridge CB3 0HA, UK.}
\altaffiltext{2}{Osservatorio Astrofisico di Arcetri, Largo E. 
Fermi 5, 50125 Firenze, Italy.}
\altaffiltext{3}{Department of Astronomy and Astrophysics, University of
California, Santa Cruz, CA 95064, USA.}

\begin{abstract}
Recent observations of high-redshift galaxies and quasars indicate that the 
hydrogen component of the intergalactic medium (IGM) must have been reionized
at some redshift $z\gta 6$. Prior to complete reionization, sources of  
ultraviolet radiation will be seen behind intervening gas that is still
neutral, and their spectra should show the red damping wing of the 
Gunn-Peterson trough. While this characteristic feature may, in principle, 
totally suppress
the \Lya emission line in the spectra of the first generation of objects in the 
universe, we show here
that the IGM in the vicinity of luminous quasars will be highly photoionized on 
several Mpc scales due to the source emission of Lyman-continuum photons. If 
the quasar lifetime
is shorter than the expansion and gas recombination timescales, the volume
ionized will be proportional to the total number of photons produced 
above 13.6 eV: the effect of this local photoionization is to greatly 
reduce the scattering opacity between the redshift of the quasar
and the boundary of its \HII region. We find that the transmission on the red side 
of the \Lya resonance is always greater than 50\% for sources radiating a total of 
$\gta 10^{69.5}$ ionizing photons into the IGM. The detection of a strong \Lya 
emission line in the spectra of bright QSOs shining for $\gta 10^7\,$yr 
cannot then be used, by itself, as a constraint on the reionization epoch. The
first signs of an object radiating prior to the transition from a neutral to 
an ionized universe may be best searched for in the spectra of luminous sources 
with a small escape fraction of Lyman-continuum photons into the IGM, or sources 
with a short duty cycle.
\end{abstract}
 
\section{Introduction}
Popular cosmological models predict that most of the intergalactic hydrogen
was reionized by the first generation of stars or accreting black holes in the
universe at a redshift
$7\lta z_\rei\lta 15$ (e.g. Ciardi \etal 2000; Haiman \& Loeb 1997; Gnedin \& 
Ostriker 1997). Studying the epoch of reheating which marks the end of 
the cosmic ``dark ages'' is crucial for determining its impact on several key 
cosmological 
issues, from the role reionization plays in allowing pregalactic objects to 
cool and make stars, to determining the small-scale structure in the temperature
fluctuations of the cosmic microwave background. 
Because of scattering off the line-of-sight due to the diffuse neutral  
intergalactic medium (IGM), the spectrum of a source at $z_\em>z_\rei$ should 
show a  Gunn-Peterson (1965, hereafter GP) absorption trough at wavelengths 
shorter 
than the local \Lya resonance, $\lambda_\obs<\lambda_\alpha(1+z_\em)$, where
$\lambda_\alpha=c/\nu_\alpha=1216\,$\AA.
Unfortunately, while actively being searched for, the first signs of an object 
radiating prior to the reionization epoch are far from unambiguous. This is 
because: (1) line blanketing from discrete \Lya forest absorbers becomes really 
severe at $z\gta 5.5$, with less than 10\% of the quasar unabsorbed continuum 
leaking through between rest-frame \Lya and \Lyb (Fan \etal 2000a; Stern \etal 
2000); and (2) the GP optical depth for resonant scattering encountered by photons 
propagating through  smoothly distributed neutral hydrogen of density 
$n_\nHI(z)$,
\beq
\tau^\b_{\rm GP}(z)=\tau_0\,{n_\nHI\over \overline n_\nH},~~~~~~~~~~~~~~~~ 
\tau_0(z)\equiv {\pi e^2 f \lambda_\alpha\over m_ec H(z)}~\overline n_\nH \simeq 
1.5\times 10^5 h^{-1} \Omega_M^{-1/2}\left(\frac {\Omega_b h^2} {0.019} \right)
\,\left({1+z\over 8}\right)^{3/2}, \label{eq:GP}
\eeq
is extremely high. Here $f$ is the oscillator strength, $e$ and $m_e$ are
the electron charge and mass, $\overline n_\nH$ is the mean density of 
hydrogen nuclei at redshift $z$, $H_0=100\,h\,\kmsmpc$ is the present-day Hubble 
constant, and ($\Omega_M, \Omega_b$) are the total matter and baryonic density 
parameter, respectively. The expression
above gives the opacity seen by any photon emitted from a source at $z_\em$ on 
the blue 
side of the \Lya line, $\lambda_\em<\lambda_\alpha$, as it is redshifted 
through the local \Lya resonance at $(1+z)=(1+z_\em)\lambda_\em/
\lambda_\alpha$. Equation (\ref{eq:GP}) shows that the 
transmitted quasar flux shortward of \Lya would be reduced to undetectable 
levels even if 99\% 
of all the cosmic baryons were to fragment at these early epochs
into discrete, mildly overdense structures, with 
only 1\% remaining in a diffuse component that was 99\% ionized. As such, the 
detection of a GP trough would not uniquely establish that   
an object is being observed prior to the reionization epoch (except, perhaps, in the case where reionization occurs extremely rapidly and the GP trough 
splits into individual Lyman series troughs for a source located at 
$(1+z_\rei)<(1+z_\em)<32(1+z_\rei)/ 27$, Haiman \& Loeb 1999). It has been 
pointed out by Miralda-Escud\'{e} (1998), however, that the rest-frame 
ultraviolet spectra of sources observed prior to complete reionization 
should show the red 
damping wing of the GP trough, as they will be seen behind a large 
column density of intervening gas that is still neutral. At $z\gta 6$, this
characteristic feature extends for more than $1500\,\kms$ to the red of the 
resonance, significantly suppressing the \Lya emission line. Measuring the 
shape of the absorption 
profile of the damping wing could provide a determination of the density of the 
neutral IGM near the source. 

In this {\it Letter} we focus on the width of the red damping 
wing -- related to the expected strength of the \Lya emission line -- in 
the spectra of very distant quasars 
as a flag of the observation of the IGM before reionization. We discuss, in
particular, the 
impact of the photoionized, Mpc-size regions which will surround individual 
luminous sources of UV radiation on the transmission of photons redward of 
the \Lya resonance, and show that the damping wing of the GP trough may nearly
completely disappear because of the lack of neutral hydrogen in the vicinity of a 
bright object. The absence in the spectra of luminous quasars of a red damping 
wing with the predicted absorption profile will not provide then unambiguous 
evidence of the observation of the IGM after cosmological reionization. 

\section{The red damping wing}

We generalize here the calculation of the absorption profile of the damping 
wing of the GP trough (Miralda-Escud\'{e} 1998) to the case  where the IGM in 
the vicinity of an object at $z_\em$ is photoionized due to the source 
emission of UV photons. When the column density of absorbing atoms is 
sufficiently large, the 
width of an absorption line substantially exceeds the value corresponding to the
dispersion of particle velocities along the line of sight. In this case, the
scattering cross-section is determined by the natural width of the \Lya 
resonance,
\beq
\sigma_\alpha(\nu)={\pi e^2 f\over m_ec}~ {\Lambda (\nu/\nu_\alpha)^4
\over 4\pi^2(\nu-\nu_\alpha)^2+ \Lambda^2(\nu/\nu_\alpha)^6/4}
\label{eq:cs}
\eeq 
(Peebles 1993), where $\Lambda=(8\pi^2 e^2 f)/(3m_ec\lambda_\alpha^2)=6.25
\times 10^8\,$ s$^{-1}$ is the rate of spontaneous radiative decay from the 
$2p$ to $1s$ energy level. The regions of the line profile for which equation 
(\ref{eq:cs}) is valid are known as ``radiation damping'' wings of the 
line. Equation (\ref{eq:cs}) 
is unapplicable close to line center, but if damping 
wings are present the transmitted flux will be essentially zero in the core 
region anyway. We assume the IGM has a constant \HI comoving density 
$n_\nHI(0)$ at all redshifts $z_\rei<z<z_i\le z_\em$, and is highly 
photoionized between the redshift of the source $z_\em$ and the boundary of 
its \HII region at $z_i$. 

The scattering optical depth at the observed wavelength $\lambda_\obs>
\lambda_\alpha(1+z_\em)$ is
\beq
\tau^\r_{\rm GP}(\lambda_\obs)=\int^{z_i}_{z_\rei} dz~{d\ell\over dz}~n_\nHI(0)
(1+z)^3~ \sigma_\alpha\left[\nu={c(1+z)\over \lambda_\obs}\right], \label{eq:tau}
\eeq 
where $d\ell/dz=c[(1+z)H(z)]^{-1}$ is the proper cosmological line element.
In an EdS universe, equation (\ref{eq:tau}) can be rewritten as
\beq
\tau^\r_{\rm GP}(\lambda_\obs)={R\,\tau_0(z_\em)\over \pi}~ \left[{\lambda_\alpha
(1+z_\em)\over \lambda_\obs}\right]^{3/2}~\int_{x_\rei}^{x_i} {dx x^{9/2}
\over  (1-x)^2 +R^2x^6},
\eeq 
where $R\equiv \Lambda\lambda_\alpha/(4\pi c)=2.02\times
10^{-8}$, $x_\rei=
(1+z_\rei)\lambda_\alpha/\lambda_\obs$, and $x_i=(1+z_i)\lambda_\alpha/
\lambda_\obs$. Far from line center (i.e. when $\parallel 1-x\parallel \gg
Rx^3$), this integral has an analytic solution (Miralda-Escud\'{e} 1998).
Figure 1 (solid curve) shows the red damping wing in the spectrum of a source at 
$z_\em=7$ assuming $\tau_0(z_\em)=3\times 10^5$, $z_\rei=6$, and $z_i=z_\em$,    
i.e. in the case the \HII region surrounding the radiation object is very 
small (because, e.g., the Lyman-continuum photons produced cannot
escape from the dense sites of star formation into the intergalactic space).
 
Equation (\ref{eq:tau}) also gives the opacity at wavelengths $\lambda_\alpha(1+z_i)
<\lambda_\obs <\lambda_\alpha(1+z_\em)$, i.e. on the blue side of the 
quasar \Lya 
emission line, due to the damping wing of the fully neutral gas along the 
line of sight at $z_\rei<z<z_i$. We will see in the next section that this 
must be augmented by the scattering optical depth of the residual \HI in 
the vicinity of the source ($z_i<z<z_\em$).

\section{Cosmological \bHII regions around isolated sources}

We now assess in details the impact of a local \HII region on the shape of the 
damping wing
profile. When an isolated point source of ionizing radiation turns on, the 
volume of ionized IGM initially grows in size at a rate fixed by the emission 
of UV photons, and an ionization front separating the \HII and \HI regions 
propagates
into the neutral gas. Most photons travel freely in the ionized bubble, and are
absorbed in a transition layer (the ``I-front''), across which the degree of
ionization changes sharply on a distance which is small compared to the 
radius of the ionized zone (this is true even in the case of a 
QSO with a hard spectrum, see Madau \& Meiksin 1991). The evolution of an 
expanding cosmological \HII region is governed by the equation
\begin{equation}
{dV_I\over dt}-3HV_I={\dot N_i\over \overline{n}_\nH}-{V_I\over
{t}_{\rm rec}}, \label{eq:dVdt}
\end{equation}
(Shapiro \& Giroux 1987; Madau, Haardt, \& Rees 1999), where $V_I$ is the 
proper photoionized volume, $\dot N_i$ is the number of H-ionizing 
photons emitted by the central source per unit time that escape into the IGM, 
\begin{equation}
{t}_{\rm rec}=(1.17\,\overline{n}_p\alpha_B\,C)^{-1}\simeq 1.3\, {\rm Gyr}
\left(\frac{\Omega_B h^2}{0.019}\right)^{-1}\left(\frac{1+z}{8}\right)^{-3}
~C^{-1}, \label{eq:trec}
\end{equation}
is the volume-averaged recombination time, $\alpha_B$ is the radiative 
recombination 
coefficient to the excited states of hydrogen (at an assumed gas temperature of 
$10^4\,$K), and the factor $C\equiv \langle n_p^2\rangle/\overline{n}_p^2>1$ 
takes 
into account the degree of clumpiness of the photoionized region, with 
$\overline{n}_p$ the mean proton density. When the source lifetime 
$t_s$ is much less than $(t_\rec,H^{-1})$,
as expected for example in the case of a quasar shining for an Eddington time
scale, $t_s=t_E=4\times 10^7\,(\epsilon/0.1)\,$yr with $\epsilon$ the radiative 
accretion efficiency, recombinations can be neglected (this would be true 
even in 
the case of a clumpy medium with $C\lta 10$ on Mpc scales, cf eqs. \ref{eq:dVdt}
and \ref{eq:trec}) and 
the evolution of the \HII region 
can be decoupled from the expansion of the universe. The volume, 
\begin{equation}
V_I\simeq {\dot N_i t_s\over \overline{n}_\nH}~~~~~~~~~~~~~~~~~~(t_s\ll 
t_\rec,H^{-1}), \label{eq:V}
\end{equation}
that is actually ionized becomes then proportional to the total number of 
Lyman-continuum photons emitted over the source lifetime. 
Note that the I-front initially expands at velocities that are close to the 
speed of light, i.e. with $r_I(t)=[3V_I(t)/4\pi]^{1/3}\simeq ct$.
In the case of very luminous, short-lived objects this `relativistic' phase may 
last a considerable fraction of the source lifetime. Nevertheless, the 
apparent size of the \HII region `seen' by \Lya photons propagating along 
the line of sight will always be
given by equation (\ref{eq:V}), as these can catch up with the I-front only 
when it slows down to subluminal velocities. 

The known QSOs at $z_\em \gta 5$ all have
ionizing luminosities that can be estimated to lie in the range $10^{56}\lta \dot 
N_i\lta 10^{58}\,$s$^{-1}$, the brightest of them being the recently discovered 
$z_\em=5.8$ quasar from the Sloan Digital Sky Survey (SDSS, Fan \etal 2000a).
Over a lifetime of (say) $10^7\,$
yr they would radiate of order $N_i=10^{70.5}-10^{72.5}$ photons above 13.6 eV.
Placed at $z_\em=7$, sources of similar power would ionize the surrounding IGM 
out to a proper distance $r_I=1.5-7\,$Mpc, corresponding to a 
Hubble expansion velocity of $\Delta v=Hr_I=3300-15800\,h\,\Omega_M^{1/2}\,\kms$, 
or to a redshift
difference between the QSO and the boundary of its \HII zone of $\Delta 
z=3.33\times 10^{-4}\,h\,\Omega_M^{1/2}(1+z_\em)^{5/2}(r_I/$Mpc$)=0.088-0.42\,h
\Omega_M^{1/2}$. The 
effect of these individual \HII regions on the flux transmission redward of the \Lya
line is shown in Figure 1. The width of the damping wing measures the column
density along the line of sight of the fraction of the IGM that is still neutral.   
The wing nearly completely disappears in the 
case of a luminous quasar, as there is very little neutral gas in its vicinity. In 
particular, the transmission is always greater than 50\% for $N_i>10^{69.5}$
photons. 

On the blue side of \Lya, in the vicinity of the source, the scattering opacity 
has two contributions, one due to the damping wing of the fully neutral gas beyond
the \HII region, and one due to the ordinary GP trough associated with   
the residual \HI in the photoionized zone.
Assuming photoionization equilibrium (this is justified for the highly ionized 
IGM and the source lifetimes considered here), at every point within the 
\HII bubble the density of neutral hydrogen is given by  
\beq
n_\nHI= {\overline n_p\over t_\rec}\left[\int_{\nu_L}^\infty {F_\nu 
\sigma_\nH(\nu) \over h_P\nu} d\nu\right]^{-1}, \label{eq:nh}
\eeq
where $h_P$ is the Planck's constant, $F_\nu$ is the incident ionizing flux 
per unit frequency -- to a first approximation 
simply the radiation emitted by the quasar reduced by geometrical dilution --
and the hydrogen photoionization cross-section (by photons above the threshold 
$h_P\nu_L=13.6\,$eV) is 
\beq
\sigma_\nH(\nu)\simeq \sigma_L\,(\nu/\nu_L)^{-3},~~~~~~~~~~~~~~~~~
\sigma_L=6.3\times 10^{-18}\,{\rm cm^2}. \label{eq:pcs}
\eeq 
For a power-law spectrum of the form $F_\nu\propto \nu^{-\alpha}$ near the hydrogen
Lyman edge, the
neutral hydrogen column through an (approximately isothermal) \HII region can 
thus be written as  
\beq
N_\nHI=\int_0^{r_I}\,dr~n_\nHI={3+\alpha\over \alpha}\,\sigma_L^{-1}\,
{t_s\over t_\rec}, \label{eq:col}
\eeq
independently of the source luminosity. Taking $\alpha=0.5$, $t_s=10^7\,$yr, and 
$t_\rec=1.3\,$Gyr at $z_\em=7$, one derives $N_\nHI=8.6\times 10^{15}\,\cmm$. 
This is too small a column (even in the case of a fast recombining clumpy medium) 
for the natural width of the line to exceed the thermal broadening. On the other 
hand, it is straightforward to derive from equations (\ref{eq:GP}), (\ref{eq:nh}),
and (\ref{eq:pcs}) 
that the GP optical depth on the blue side of \Lya due to partially ionized gas
at a distance $r<r_I$ from the QSO is given by
\beq
\tau^\b_{\rm GP}(r)\simeq 2\,
\left({r\over r_I}\right)^2\, \left({t_s\over t_\rec}\right)\,
\left({cH^{-1}\over r_I}\right)\,
\left({3+\alpha\over \alpha}\right).  \label{eq:th2}
\eeq
While the first two terms are of course smaller than unity, the third term is
rather large, $cH^{-1}/r_I\simeq 130\,h^{-1}\Omega_M^{-1/2}/r_I\,$Mpc at $z_\em=7$,
and this means that only the inner parts of the \HII region will be
optically thin to the classical GP absorption. Note that $\tau^\b_{\rm GP}$
just depends on the quasar luminosity, not on $t_s$ and $r_I$ separately, and 
that equations
(\ref{eq:nh}), (\ref{eq:col}), and (\ref{eq:th2}) assume: (1) a steady luminosity,
and (2) a uniform medium at the mean background density. 
The hydrogen neutral fraction will actually depend on the mean ionizing flux 
over the last $t_\rec (n_\nHI/\overline n_p)\ll t_\rec$ yr, while halos having 
collapsed from (say) 3$\sigma$ fluctuations may actually be sitting in slightly 
overdense regions.

\section{Summary}

The lack of a GP trough -- i.e. the detection of transmitted flux shortward of 
the \lya wavelength -- observed in the spectrum of the $z_\em=5.8$ SDSS quasar 
(Fan \etal 2000a) indicates that the 
IGM was already highly ionized at that redshift, and that sources of ultraviolet
photons were present in significant numbers when the universe was less 
than 6\% of its current age. At the time of writing, 
three quasars have already been found in the range $5\lta z_\em\lta 5.5$
(Zheng \etal 2000; Stern \etal 2000; Fan \etal 2000b), and two galaxies 
have been spectroscopically confirmed at $z_\em\gta 5.6$ (Hu, McMahon, \& Cowie
1999; Weymann \etal 1998). It is estimated 
that the SDSS could reveal tens of additional sources at these epochs, and one 
QSO at $z_\em\gta 6$ about twice as luminous as 3C273 in every 1500 deg$^2$ of the 
survey (Fan \etal 2000a). 

Near-future studies of the rest-frame UV spectra of high-redshift quasars and 
star-forming galaxies could then conceivably provide important probes of the 
formative early stages of cosmic evolution. In this {\it Letter} we have discussed 
one of the first signs of an object radiating prior to the transition from a neutral 
to an ionized universe, the red damping wing of the GP trough. We have shown that
the local photoionized zones which will inevitably surround luminous quasars at 
early epochs will greatly reduce the scattering opacity between the redshift of 
the source and the boundary of its Mpc-size \HII region, thus increasing the 
transmission of photons on the red side of the \Lya resonance. 
To better gauge this effect on real data, we have plotted in Figure 2 the Keck/LRIS
spectrum of the faint $z_\em=5.5$ quasar RD J0301117$+$002025 (Stern \etal 2000), 
redshifted to $z_\em=7$. The figure depicts the same $800\,$\AA-wide region of the 
observed spectrum around the \Lya resonance, together with the transmission 
$\exp(-\tau_{\rm GP})$ assuming the QSO is being observed prior to the reionization 
epoch at $z_\rei=6$. This should be taken just as an illustrative example, as
in some numerical simulations (e.g. Ciardi \etal 2000) reionization was
already well in progress prior to redshift 6, and the form of the damping 
profile would be different in the case of patchy ionization along the line of
sight. 
Four cases are shown, as the emission rate of UV 
photons which ionize the IGM in the vicinity of the quasar is increased from $\dot 
N_i=0$ to $\dot N_i=10^{56}, 10^{57},$ and $10^{58}$ s$^{-1}$. The 
calculations assume a source lifetime of $t_s=10^7\,$yr. 
The suppression effect of the \Lya emission line by the wing of 
the GP trough, while clearly visible in the absence of a local \HII region (top-left
panel), weakens significantly as the size of the photoionized zone increases.
On the blue side of the resonance, the transmission profile is due to the 
combination of two effects: (1) the damping wing -- i.e. the curve in the top-left
panel at $\lambda_\obs>\lambda_\alpha(1+z_\em)$ -- shifted to the blue by the
difference between the redshift of the \HII region boundary and the quasar redshift;
and (2) the ordinary GP optical depth from the residual neutral gas in the \HII
zone. The first of these contributions depends on $N_i$ (i.e. on the luminosity
$\dot N_i$ multiplied by the source lifetime $t_s$), whereas the second just   
depends on $\dot N_i$.
Figure 3 shows a similar set  of spectra and transmissions at earlier epochs, 
with $z_\em=9$, $z_\rei=8$, and a
recombination timescale which is $(10/8)^3$ times shorter. 

Depending on the amount of H-ionizing photons which escape the \HI 
galactic layers into the intergalactic space, the red damping wing 
may be more easily detected in the spectra of star-forming 
early galaxies (as well as luminous sources with a short duty cycle). 
For standard initial mass functions, about 3000--4000 photons are produced 
above 13.6 eV 
for every stellar baryon. Thus a galaxy which turns a gas mass $M_*$ into
stars will radiate a total of about $10^{68.6}\,(M_*/10^{8}\,\msun)$ 
Lyman-continuum photons.
It is clear from Figure 1 that the effect of its local \HII region on the shape of 
the GP absorption profile would be negligible if the escape fraction was 
(say) $\lta 10\%$. In this case, large \HII regions
may still be expected if these early galaxies were highly clustered on Mpc
scales.  

\acknowledgments
Support for this work was provided by NASA through ATP grant NAG5--4236 (P. M.),
by a B. Rossi Visiting Fellowship at the Observatory of Arcetri (P. M.), and by 
the Royal Society (M. J. R.). We are indebted to D. Stern and H. Spinrad for 
providing the spectrum of quasar RD J030117$+$002025. Results similar to the
ones presented here have been reached independently by Cen \& Haiman (2000).

\begin{figure}
\plotone{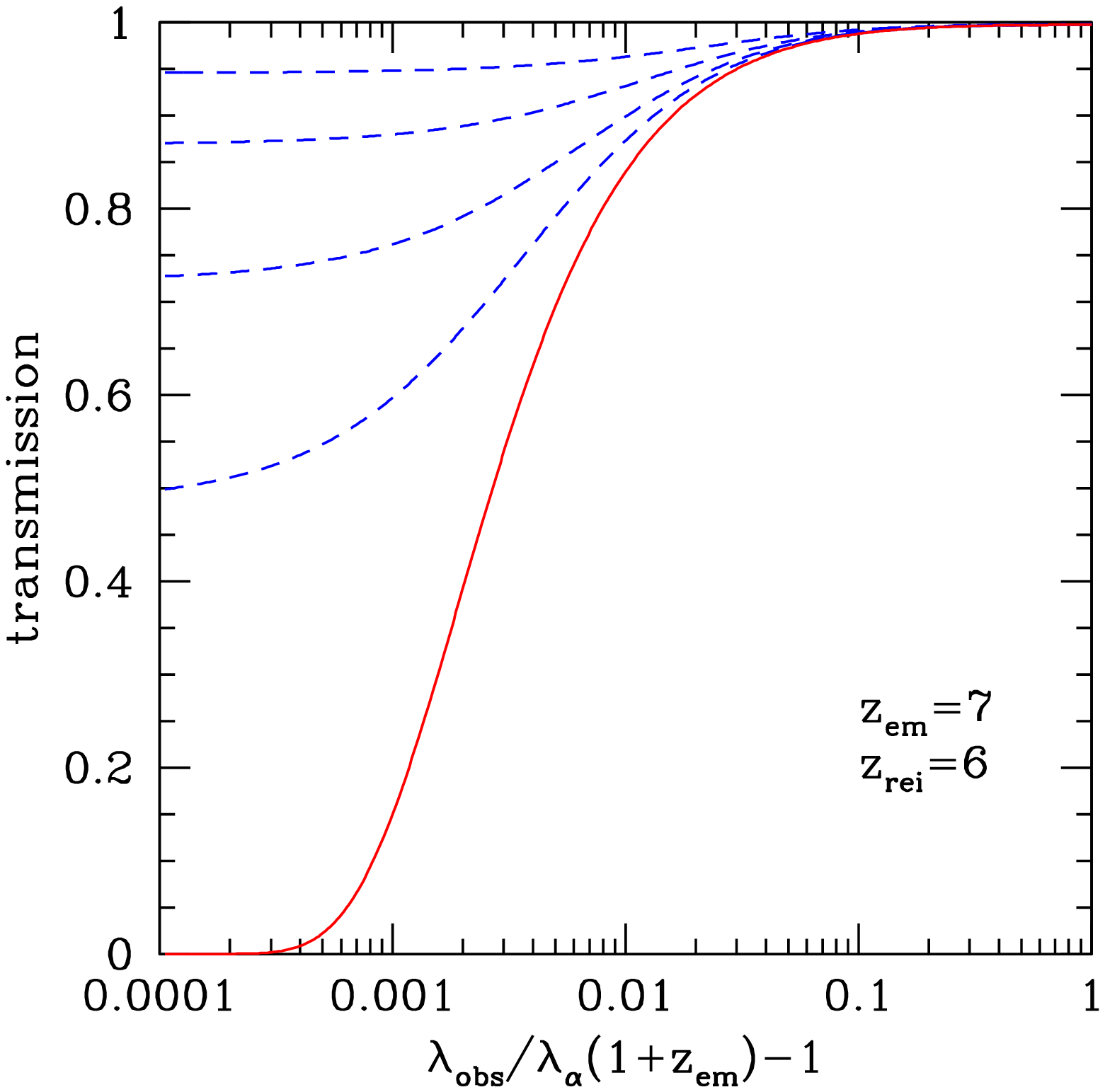}
\caption{The red damping wing of the Gunn-Peterson trough for a source at $z_\em=7$.
The transmission $\exp(-\tau^\r_{\rm GP})$ is plotted as a function of the 
fractional wavelength interval from the \Lya resonance at $\lambda_\alpha(1+z_\em)$. 
The neutral IGM has $\tau_0=3\times 10^5$ at $z_\em$ (cf. eq. 1), and is 
assumed to be completely reionized by $z_\rei=6$.
{\it Solid curve:} absorption profile neglecting the effect of the local \HII
region around the source ($N_i=0$). Note how a fraction $\gta \exp(-0.5)=0.6$ of 
the radiated flux will only be transmitted $\gta 1200\,\kms$ (corresponding to 
$\gta 40\,$\AA) to the red of the resonance. 
{\it Dashed curves:} absorption profiles including the local 
\HII zone generated by a QSO shining for $10^7\,$yr. From top to bottom, the
curves are plotted for four decreasing ionizing luminosities, 
$\dot N_i=10^{58}$, $10^{57},$ $10^{56}\,$, and $10^{55}\,$s$^{-1}$ 
(Einstein-de Sitter universe with $h=0.5$). 
\label{fig1}}
\vspace{+0.5cm}
\end{figure}

\begin{figure}
\plotone{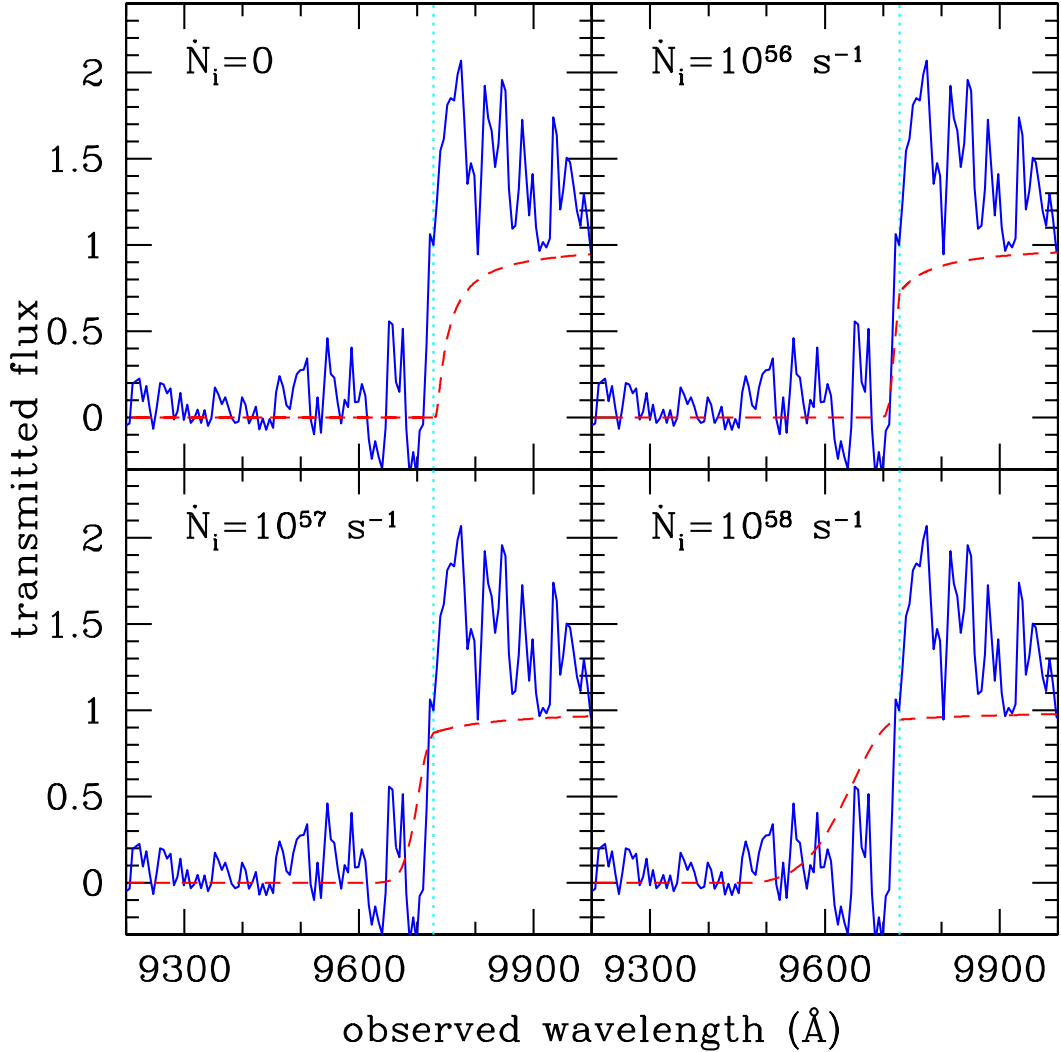}
\caption{The Keck/LRIS spectrum of the faint $z_\em=5.5$ quasar RD 
J0301117$+$002025 (Stern \etal 2000), redshifted to $z_\em=7$. The dotted 
vertical lines indicate the expected wavelength of the \Lya resonance. Fluxes
have arbitrary normalizations. The dashed curves show the transmission 
$\exp(-\tau_{\rm GP})$ through a uniform IGM, assuming the QSO is being 
observed prior to the reionization epoch at $z_\rei=6$. Four different cases are 
shown, as the rate of emission of Lyman-continuum photons which ionize the IGM in 
the vicinity of the source is varied from 
$\dot N_i=0$ to $\dot N_i=10^{56}, 10^{57},$ and $10^{58}$ s$^{-1}$. 
The calculations assume a quasar lifetime of $t_s=10^7\,$yr, a power-law 
spectrum of the form    
$F_\nu\propto \nu^{-\alpha}$ with $\alpha=0.5$ near the hydrogen Lyman edge, 
a recombination timescale of $t_{\rm rec}=1.3\,$Gyr, and
an Einstein-de Sitter universe with $h=0.5$.
\label{fig2}}
\vspace{+0.5cm}
\end{figure}

\begin{figure}
\plotone{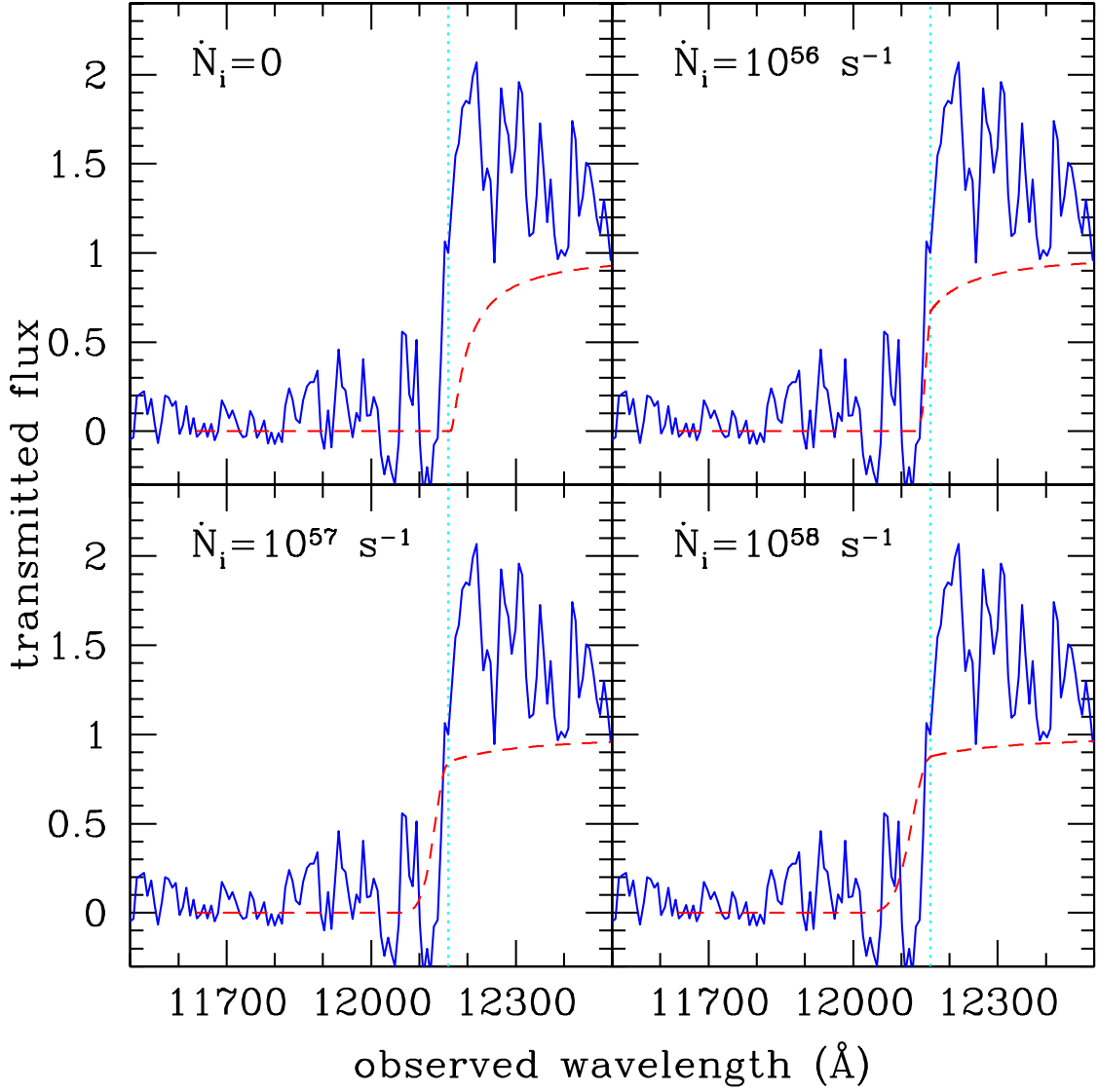}
\caption{Same as Figure 2, but with $z_\em=9$, $z_\rei=8$, and a recombination 
timescale $(10/8)^3$ times shorter. 
\label{fig3}}
\vspace{+0.5cm}
\end{figure}

\end{document}